# On separable Fokker–Planck equations with a constant diagonal diffusion matrix

Alexander Zhalij*

Institute of Mathematics of the Academy of Sciences of Ukraine,

Tereshchenkivska Street 3, 252004 Kyiv, Ukraine

### Abstract

We classify (1+3)-dimensional Fokker–Planck equations with a constant diagonal diffusion matrix that are solvable by the method of separation of variables. As a result, we get possible forms of the drift coefficients $B_1(\vec{x}), B_2(\vec{x}), B_3(\vec{x})$ providing separability of the corresponding Fokker–Planck equations and carry out variable separation in the latter. It is established, in particular, that the necessary condition for the Fokker–Planck equation to be separable is that the drift coefficients $\vec{B}(\vec{x})$ must be linear. We also find the necessary condition for R–separability of the Fokker–Planck equation. Furthermore, exact solutions of the Fokker–Planck equation with separated variables are constructed.

## I. Introduction

The diffusion processes play an important role in different fields of physics, chemistry and biology [1, 2] and have very broad applications in technics. For example, they play a decisive role in electronics [1]. That is why, analytical solutions of diffusion equations were, are and will be of great importance for applications, since they provide new insights into the nature of the diffusion processes described by the equations in question. As an example, lets us mention the famous Black–Scholes models, whose success is based, in particular, on an analytical solution constructed in an explicit form.

*e-mail: zhaliy@imath.kiev.ua

Unfortunately, the diffusion (Fokker-Planck) equations, that are used in applications, have variable coefficients and cannot be integrated by the standard Fourier transform method. In fact, the only available efficient way for constructing analytical solutions of partial differential equations (PDEs) with variable coefficients is the method of separation of variables.

The simplest and most widely used in applications is the case of constant diffusion matrix. Therefore the principal object of the study in the present paper is a problem of separation of variables in the Fokker–Planck equation (FPE) [3] with a constant diagonal diffusion matrix

$$u_t + \Delta u + (B_a(\vec{x})u)_{x_a} = 0, \qquad (1)$$

where $\vec{B}(\vec{x}) = (B_1(\vec{x}), B_2(\vec{x}), B_3(\vec{x}))$ is the drift velocity vector. Here $u = u(t, \vec{x})$ and $B_i(\vec{x})$, $i = 1, 2, 3$ are smooth real–valued functions. Hereafter, the subscript $x_a$ implies partial differentiation and, moreover, summation over the repeated Latin indices from 1 to 3 is understood. The matrix of the constant diffusion coefficients is reduced to the unit matrix by proper simple transformation.

In the present paper we solve the problem of variable separation in FPE (1) into second-order ordinary differential equations in a sense that we obtain possible forms of the drift coefficients $B_1(\vec{x}), B_2(\vec{x}), B_3(\vec{x})$ providing separability of (1). Furthermore, we construct inequivalent coordinate systems enabling to separate variables in the corresponding FPEs and carry out variable separation.

The separability criteria for the one-dimensional FPE have been obtained in [4].

The problem of variable separation in the three-dimensional FPE was considered in a restricted sense by Sukhomlin in [5]. He used the symmetry approach which is based on the well-known fact that a solution with separated variables is a common eigenfunction of three first- or second-order differential operators which commute with each other and with the operator of the equation under consideration (see, [6] and the references therein). Sukhomlin has obtained some drift coefficients providing separability of FPE

(1) and carried out separation of variables in the latter. But his results are far from being complete and systematic.

Our analysis is based on the direct approach to variable separation in linear PDEs suggested in [7]–[9]. It has been successfully applied to solving variable separation problem in the wave [7] and Schrödinger equations [8]–[12] with variable coefficients.

## II. Separation of variables in the Fokker–Planck equation

Let us formulate briefly the algorithm of variable separation in FPE (1) following [9].

We say that FPE (1) is separable in a coordinate system $t, \omega_a = \omega_a(t, \vec{x})$, $a = 1, 2, 3$ if the separation Ansatz

$$u(t, \vec{x}) = \varphi_0(t) \prod_{a=1}^{3} \varphi_a\left(\omega_a(t, \vec{x}), \vec{\lambda}\right) \qquad (2)$$

reduces PDE (1) to four ordinary differential equations for the functions $\varphi_\mu$, $(\mu = 0, 1, 2, 3)$

$$\varphi_0' = U_0(t, \varphi_0; \vec{\lambda}), \quad \varphi_a'' = U_a(\omega_a, \varphi_a, \varphi_a'; \vec{\lambda}). \qquad (3)$$

Here $U_0, \ldots, U_3$ are some smooth functions of the indicated variables, $\vec{\lambda} = (\lambda_1, \lambda_2, \lambda_3) \in \Lambda = \{$an open domain in $\mathbf{R}^3\}$ are separation constants (spectral parameters, eigenvalues) and, what is more,

$$\text{rank} \left\| \frac{\partial U_\mu}{\partial \lambda_a} \right\|_{\mu=0\ a=1}^{3\ \ 3} = 3. \qquad (4)$$

The above condition secures essential dependence of a solution with separated variables on the separation constants $\vec{\lambda}$.

The principal steps of the procedure of variable separation in FPE (1) are as follows

1. We insert the Ansatz (2) into FPE and express the derivatives $\varphi_0'$, $\varphi_1''$, $\varphi_2''$, $\varphi_3''$ in terms of functions $\varphi_0$, $\varphi_1$, $\varphi_2$, $\varphi_3$, $\varphi_1'$, $\varphi_2'$, $\varphi_3'$ using equations (3).

2. We regard $\varphi_0$, $\varphi_1$, $\varphi_2$, $\varphi_3$, $\varphi_1'$, $\varphi_2'$, $\varphi_3'$, $\lambda_1$, $\lambda_2$, $\lambda_3$ as the new independent variables $y_1,\ldots, y_{10}$. As the functions $\omega_1$, $\omega_2$, $\omega_3$ are independent of the variables $y_1,\ldots, y_{10}$, we can split by these variables and get an over-determined system of nonlinear partial differential equations for unknown functions $\omega_1$, $\omega_2$, $\omega_3$.

3. After solving the above system we get an exhaustive description of coordinate systems providing separability of FPE.

Having performed the first two steps of the above algorithm we arrive at the conclusion that the separation equations (3) are linear both in $\varphi_0,\ldots, \varphi_3$ and $\lambda_1, \lambda_2, \lambda_3$.

Next, we introduce an equivalence relation $\mathcal{E}$ on the set of all coordinate systems providing separability of FPE. We say that two coordinate systems $t, \omega_1, \omega_2, \omega_3$ and $\tilde{t}, \tilde{\omega}_1, \tilde{\omega}_2, \tilde{\omega}_3$ are *equivalent* if the corresponding Ansatzes (2) are transformed one into another by the invertible transformations of the form

$$t \to \tilde{t} = f_0(t), \quad \omega_i \to \tilde{\omega}_i = f_i(\omega_i), \tag{5}$$

where $f_0,\ldots, f_3$ are some smooth functions and $i = 1, 2, 3$. These equivalent coordinate systems give rise to the same solution with separated variables, therefore we shall not distinguish between them. The equivalence relation (5) splits the set of all possible coordinate systems into equivalence classes. In a sequel, when presenting the lists of coordinate systems enabling us to separate variables in FPE we will give only one representative for each equivalence class.

Following [12] we choose the reduced equations (3) to be

$$\varphi_0' = (T_0(t) - T_i(t)\lambda_i)\varphi_0, \quad \varphi_a'' = (F_{a0}(\omega_a) + F_{ai}(\omega_a)\lambda_i)\varphi_a, \tag{6}$$

where $T_0, T_i, F_{a0}, F_{ai}$ are some smooth functions of the indicated variables, $a = 1, 2, 3$. With this remark the system of nonlinear PDEs for unknown functions $\omega_1$, $\omega_2$, $\omega_3$ takes the form

$$\frac{\partial \omega_i}{\partial x_a}\frac{\partial \omega_j}{\partial x_a} = 0, \quad i \neq j, \quad i, j = 1, 2, 3; \tag{7}$$

$$\sum_{i=1}^{3} F_{ia}(\omega_i)\frac{\partial \omega_i}{\partial x_j}\frac{\partial \omega_i}{\partial x_j} = T_a(t), \quad a = 1, 2, 3; \tag{8}$$

$$B_j \frac{\partial \omega_a}{\partial x_j} + \frac{\partial \omega_a}{\partial t} + \Delta \omega_a = 0, \quad a = 1, 2, 3; \tag{9}$$

$$\sum_{i=1}^{3} F_{i0}(\omega_i) \frac{\partial \omega_i}{\partial x_j} \frac{\partial \omega_i}{\partial x_j} + T_0(t) + \frac{\partial B_a}{\partial x_a} = 0. \tag{10}$$

Thus the problem of variable separation in FPE reduces to integrating system of ten *nonlinear* PDEs for three functions. What is more, some coefficients are arbitrary functions, which should be determined while integrating equations (7)–(10). We have succeeded in constructing their general solution which yields, in particular, all possible functions $B_1(\vec{x}), B_2(\vec{x}), B_3(\vec{x})$ such that FPE (1) is solvable by the method of separation of variables.

The system of equations (7), (8) has been integrated in [12].

**Lemma 1** *The general solution* $\vec{\omega} = \vec{\omega}(t, \vec{x})$ *of system of partial differential equations (7), (8) is given implicitly by the following formulae:*

$$\vec{x} = \mathcal{T}(t) H(t) \vec{z}(\vec{\omega}) + \vec{w}(t). \tag{11}$$

Here $\mathcal{T}(t)$ is the time-dependent $3 \times 3$ orthogonal matrix:

$$\mathcal{T}(t) = \begin{pmatrix} \cos\alpha \cos\beta - \sin\alpha \sin\beta \cos\gamma \\ \sin\alpha \cos\beta + \cos\alpha \sin\beta \cos\gamma & \rightarrow \\ \sin\beta \sin\gamma \end{pmatrix} \tag{12}$$

$$\rightarrow \begin{matrix} -\cos\alpha \sin\beta - \sin\alpha \cos\beta \cos\gamma & \sin\alpha \sin\gamma \\ -\sin\alpha \sin\beta + \cos\alpha \cos\beta \cos\gamma & -\cos\alpha \sin\gamma \\ \cos\beta \sin\gamma & \cos\gamma \end{matrix} \Bigg),$$

$\alpha, \beta, \gamma$ being arbitrary smooth functions of $t$; $\vec{z} = \vec{z}(\vec{\omega})$ is given by one of the eleven formulae

1. Cartesian coordinate system

   $z_1 = \omega_1, \quad z_2 = \omega_2, \quad z_3 = \omega_3.$

2. Cylindrical coordinate system

   $z_1 = e^{\omega_1} \cos\omega_2, \quad z_2 = e^{\omega_1} \sin\omega_2, \quad z_3 = \omega_3.$

3. Parabolic cylindrical coordinate system

   $z_1 = (\omega_1^2 - \omega_2^2)/2, \quad z_2 = \omega_1 \omega_2, \quad z_3 = \omega_3.$

4. Elliptic cylindrical coordinate system

$$z_1 = a \cosh \omega_1 \cos \omega_2, \quad z_2 = a \sinh \omega_1 \sin \omega_2, \quad z_3 = \omega_3.$$

5. Spherical coordinate system
$$z_1 = \omega_1^{-1} \operatorname{sech} \omega_2 \cos \omega_3, \quad z_2 = \omega_1^{-1} \operatorname{sech} \omega_2 \sin \omega_3,$$
$$z_3 = \omega_1^{-1} \tanh \omega_2.$$

6. Prolate spheroidal coordinate system
$$z_1 = a \operatorname{csch} \omega_1 \operatorname{sech} \omega_2 \cos \omega_3, \quad z_2 = a \operatorname{csch} \omega_1 \operatorname{sech} \omega_2 \sin \omega_3,$$
$$z_3 = a \coth \omega_1 \tanh \omega_2. \tag{13}$$

7. Oblate spheroidal coordinate system
$$z_1 = a \sec \omega_1 \operatorname{sech} \omega_2 \cos \omega_3, \quad z_2 = a \sec \omega_1 \operatorname{sech} \omega_2 \sin \omega_3,$$
$$z_3 = a \tan \omega_1 \tanh \omega_2.$$

8. Parabolic coordinate system
$$z_1 = e^{\omega_1 + \omega_2} \cos \omega_3, \quad z_2 = e^{\omega_1 + \omega_2} \sin \omega_3,$$
$$z_3 = (e^{2\omega_1} - e^{2\omega_2})/2.$$

9. Paraboloidal coordinate system
$$z_1 = 2a \cosh \omega_1 \cos \omega_2 \sinh \omega_3, \quad z_2 = 2a \sinh \omega_1 \sin \omega_2 \cosh \omega_3,$$
$$z_3 = a(\cosh 2\omega_1 + \cos 2\omega_2 - \cosh 2\omega_3)/2.$$

10. Ellipsoidal coordinate system
$$z_1 = ik^{-1}(k')^{-1} \operatorname{dn}(\omega_1, k) \operatorname{dn}(\omega_2, k) \operatorname{dn}(\omega_3, k),$$
$$z_2 = -k(k')^{-1} \operatorname{cn}(\omega_1, k) \operatorname{cn}(\omega_2, k) \operatorname{cn}(\omega_3, k),$$
$$z_3 = k \operatorname{sn}(\omega_1, k) \operatorname{sn}(\omega_2, k) \operatorname{sn}(\omega_3, k).$$

11. Conical coordinate system
$$z_1 = \omega_1^{-1}(k')^{-1} \operatorname{dn}(\omega_2, k) \operatorname{dn}(\omega_3, k),$$
$$z_2 = i\omega_1^{-1} k(k')^{-1} \operatorname{cn}(\omega_2, k) \operatorname{cn}(\omega_3, k),$$
$$z_3 = \omega_1^{-1} k \operatorname{sn}(\omega_2, k) \operatorname{sn}(\omega_3, k);$$

$H(t)$ is the $3 \times 3$ diagonal matrix

$$H(t) = \begin{pmatrix} h_1(t) & 0 & 0 \\ 0 & h_2(t) & 0 \\ 0 & 0 & h_3(t) \end{pmatrix}, \tag{14}$$

where

(a) $h_1(t), h_2(t), h_2(t)$ are arbitrary smooth functions for the completely split coordinate system (case 1 from (13)),

(b) $h_1(t) = h_2(t)$, $h_1(t), h_3(t)$ being arbitrary smooth functions, for the partially split coordinate systems (cases 2–4 from (13)),

(c) $h_1(t) = h_2(t) = h_3(t)$, $h_1(t)$ being an arbitrary smooth function, for non-split coordinate systems (cases 5–11 from (13))

and $\vec{w}(t)$ stands for the vector-column whose entries $w_1(t), w_2(t), w_3(t)$ are arbitrary smooth functions of $t$.

Here we use the usual notations for the trigonometric, hyperbolic and Jacobi elliptic functions, $k$ $(0 < k < 1)$ being the modulus of the latter and $k' = (1 - k^2)^{1/2}$. To obtain real values for $z_1, z_2, z_3$ for the ellipsoidal coordinates (system 10) we choose $\omega_1$ real, $\omega_2$ complex such that $\text{Re } \omega_2 = K$, and $\omega_3$ complex such that $\text{Im } \omega_3 = K'$, where $K, K'$ are defined by

$$K(k) = \int_0^{\pi/2} (1 - k^2 \sin^2 \theta)^{-1/2} d\theta, \quad K' = K(k').$$

To cover all real values of $z_1, z_2, z_3$ once, it is sufficient to let $\omega_1$ vary in the interval $[-K, K]$, $\omega_2$ vary in $[K - iK', K + iK']$ (parallel to imaginary axis), and $\omega_3$ vary in $[-K + iK', K + iK']$. For the conical coordinates (system 11) $\omega_1, \omega_2, \omega_3$ have the range $0 \leq \omega_1$, $-2K < \omega_2 < 2K$, $K \leq \omega_3 < K + 2iK'$. For more details about elliptic functions, see [14].

Moreover we have obtained the explicit forms of the functions $F_{ij}$, $(i, j = 1, 2, 3)$ for each class of functions $\vec{z} = \vec{z}(\vec{\omega})$ given in (13). The results are presented below in the form of $3 \times 3$ Stäckel matrices [15] $\mathcal{F}_1, \ldots, \mathcal{F}_{11}$, whose $(i, j)$th entry is the corresponding function $F_{ij}(\omega_i)$.

$$\mathcal{F}_1 = \begin{pmatrix} 1 & 0 & 0 \\ 0 & 1 & 0 \\ 0 & 0 & 1 \end{pmatrix}, \quad \mathcal{F}_2 = \begin{pmatrix} e^{2\omega_1} & -1 & 0 \\ 0 & 1 & 0 \\ 0 & 0 & 1 \end{pmatrix},$$

$$\mathcal{F}_3 = \begin{pmatrix} \omega_1^2 & -1 & 0 \\ \omega_2^2 & 1 & 0 \\ 0 & 0 & 1 \end{pmatrix}, \quad \mathcal{F}_4 = \begin{pmatrix} a^2 \cosh^2 \omega_1 & 1 & 0 \\ -a^2 \cos^2 \omega_2 & -1 & 0 \\ 0 & 0 & 1 \end{pmatrix},$$

$$\mathcal{F}_5 = \begin{pmatrix} \omega_1^{-4} & -\omega_1^{-2} & 0 \\ 0 & \cosh^{-2} \omega_2 & -1 \\ 0 & 0 & 1 \end{pmatrix},$$

$$\mathcal{F}_6 = \begin{pmatrix} a^2 \sinh^{-4} \omega_1 & -\sinh^{-2} \omega_1 & -1 \\ a^2 \cosh^{-4} \omega_2 & \cosh^{-2} \omega_2 & -1 \\ 0 & 0 & 1 \end{pmatrix},$$

$$\mathcal{F}_7 = \begin{pmatrix} a^2 \cos^{-4} \omega_1 & -\cos^{-2} \omega_1 & 1 \\ -a^2 \cosh^{-4} \omega_2 & \cosh^{-2} \omega_2 & -1 \\ 0 & 0 & 1 \end{pmatrix}, \tag{15}$$

$$\mathcal{F}_8 = \begin{pmatrix} e^{4\omega_1} & -e^{2\omega_1} & -1 \\ e^{4\omega_2} & e^{2\omega_2} & -1 \\ 0 & 0 & 1 \end{pmatrix},$$

$$\mathcal{F}_9 = \begin{pmatrix} a^2 \cosh^2 2\omega_1 & -a\cosh 2\omega_1 & -1 \\ -a^2 \cos^2 2\omega_2 & a\cos 2\omega_2 & 1 \\ a^2 \cosh^2 2\omega_3 & a\cosh 2\omega_3 & -1 \end{pmatrix},$$

$$\mathcal{F}_{10} = k^2 \begin{pmatrix} \mathrm{sn}^4(\omega_1, k) & \mathrm{sn}^2(\omega_1, k) & 1 \\ \mathrm{sn}^4(\omega_2, k) & \mathrm{sn}^2(\omega_2, k) & 1 \\ \mathrm{sn}^4(\omega_3, k) & \mathrm{sn}^2(\omega_3, k) & 1 \end{pmatrix},$$

$$\mathcal{F}_{11} = \begin{pmatrix} \omega_1^{-4} & -\omega_1^{-2} & 0 \\ 0 & -k^2 \mathrm{sn}^2(\omega_2, k) & 1 \\ 0 & -k^2 \mathrm{sn}^2(\omega_3, k) & 1 \end{pmatrix}.$$

We have also got the expressions for $T_1(t)$, $T_2(t)$, $T_3(t)$ in terms of $h_1(t)$, $h_2(t)$, $h_3(t)$:

$$\begin{aligned} 1. &\quad T_i = h_i^{-2}, \quad i = 1, 2, 3; \\ 2-4. &\quad T_1 = h_1^{-2}, \quad T_2 = 0, \quad T_3 = h_3^{-2}; \\ 5-11. &\quad T_1 = h_1^{-2}, \quad T_2 = T_3 = 0. \end{aligned} \tag{16}$$

In view of the above it is not difficult to integrate the remaining equations from the system under study.

Note that we have chosen the coordinate systems $\omega_1, \omega_2, \omega_3$ with the use of the equivalence relation $\mathcal{E}$ (5) in such a way that the relations

$$\Delta \omega_a = 0, \quad a = 1, 2, 3 \tag{17}$$

hold for all the cases 1–11 in (13). Solving (9) with respect to $B_j(\vec{x}), i = 1, 2, 3$ we get (see, also [12])

$$\vec{B}(\vec{x}) = \mathcal{M}(t)(\vec{x} - \vec{w}) + \dot{\vec{w}}. \tag{18}$$

Here we use the designation

$$\mathcal{M}(t) = \dot{\mathcal{T}}(t)\mathcal{T}^{-1}(t) + \mathcal{T}(t)\dot{H}(t)H^{-1}(t)\mathcal{T}^{-1}(t), \tag{19}$$

where $\mathcal{T}(t)$, $H(t)$ are variable $3 \times 3$ matrices defined by formulae (12) and (14), correspondingly, $\vec{w} = (w_1(t), w_2(t), w_3(t))^T$ and the dot over a symbol means differentiation with respect to $t$.

As the functions $B_1, B_2, B_3$ are independent of $t$, it follows from (18) that

$$\vec{B}(\vec{x}) = \mathcal{M}\vec{x} + \vec{v}, \quad \vec{v} = \text{const}, \tag{20}$$
$$\mathcal{M} = \text{const}, \tag{21}$$
$$\dot{\vec{w}} = \mathcal{M}\vec{w} + \vec{v}. \tag{22}$$

Taking into account that $\dot{\mathcal{T}}\mathcal{T}^{-1}$ is antisymmetric and $\mathcal{T}\dot{H}H^{-1}\mathcal{T}^{-1}$ is symmetric part of $\mathcal{M}$ (19), correspondingly, we get from (21)

$$\dot{\mathcal{T}}(t)\mathcal{T}^{-1}(t) = \text{const}, \tag{23}$$
$$\mathcal{T}(t)\dot{H}(t)H^{-1}(t)\mathcal{T}^{-1}(t) = \text{const}. \tag{24}$$

Relation (23) yields the system of three ordinary differential equations for the functions $\alpha(t), \beta(t), \gamma(t)$

$$\begin{aligned} \dot{\alpha} + \dot{\beta}\cos\gamma &= C_1, \\ \dot{\beta}\cos\alpha\sin\gamma - \dot{\gamma}\sin\alpha &= C_2, \\ \dot{\beta}\sin\alpha\sin\gamma + \dot{\gamma}\cos\alpha &= C_3, \end{aligned} \tag{25}$$

where $C_1, C_2, C_3$ are arbitrary real constants. Integrating the above system we obtain the following form of the matrix $\mathcal{T}(t)$:

$$\mathcal{T}(t) = \mathcal{C}_1 \tilde{\mathcal{T}} \mathcal{C}_2, \tag{26}$$

where $\mathcal{C}_1, \mathcal{C}_2$ are arbitrary constant $3 \times 3$ orthogonal matrices and

$$\tilde{\mathcal{T}} = \begin{pmatrix} -\cos s \cos bt & \sin s & \cos s \sin bt \\ \sin bt & 0 & \cos bt \\ \sin s \cos bt & \cos s & -\sin s \sin bt \end{pmatrix} \tag{27}$$

with arbitrary constants $b$ and $s$.

The substitution of equality (26) into (24) with subsequent differentiation of the obtained equation with respect to $t$ yields

$$\mathcal{C}_2^{-1}\tilde{\mathcal{T}}^{-1}\dot{\tilde{\mathcal{T}}}\,\mathcal{C}_2 L + \dot{L} + L\,\mathcal{C}_2^{-1}(\tilde{\mathcal{T}}^{-1})\dot{\tilde{\mathcal{T}}}\,\mathcal{C}_2 = 0, \qquad (28)$$

where $L = \dot{H}H^{-1}$, i.e. $l_i = \dot{h}_i/h_i$, $i = 1, 2, 3$. From (28) we have

$$\begin{aligned}
& l_i = \text{const}, \quad i = 1,2,3; \\
& b\,(l_1 - l_2)\cos\alpha_2\,\sin\gamma_2 = 0, \\
& b\,(l_1 - l_3)(-\sin\alpha_2\,\sin\beta_2 + \cos\alpha_2\,\cos\beta_2\,\cos\gamma_2) = 0, \\
& b\,(l_2 - l_3)(\sin\alpha_2\,\cos\beta_2 + \cos\alpha_2\,\sin\beta_2\,\cos\gamma_2) = 0,
\end{aligned} \qquad (29)$$

where $\alpha_2, \beta_2, \gamma_2$ are the Euler angles for the orthogonal matrix $\mathcal{C}_2$. Thus we obtain the following forms of $h_i$:

$$h_i = c_i \exp(l_i t), \quad c_i = \text{const}, \quad l_i = \text{const}, \quad i = 1, 2, 3. \qquad (30)$$

From (29) we get the possible forms of $b$, $l_i$ and $\mathcal{C}_2$:

$$\begin{aligned}
(i) \quad & b = 0, \quad l_1, l_2, l_3 \text{ are arbitrary constants,} \\
& \mathcal{C}_2 \text{ is an arbitrary constant orthogonal matrix;} \\
(ii) \quad & b \neq 0, \quad l_1 = l_2 = l_3, \\
& \mathcal{C}_2 \text{ is an arbitrary constant orthogonal matrix;} \\
(iii) \quad & b \neq 0, \quad l_1 = l_2 \neq l_3, \\
& \mathcal{C}_2 = \begin{pmatrix} \varepsilon_1 \cos\theta & -\varepsilon_1 \sin\theta & 0 \\ 0 & 0 & -\varepsilon_1\varepsilon_2 \\ \varepsilon_2 \sin\theta & \varepsilon_2 \cos\theta & 0 \end{pmatrix},
\end{aligned} \qquad (31)$$

where $\varepsilon_1, \varepsilon_2 = \pm 1$, and $\theta$ is arbitrary constant. We do not adduce cases $b \neq 0$, $l_1 \neq l_2 = l_3$ and $b \neq 0$, $l_2 \neq l_1 = l_3$ because they are equivalent to case $(iii)$.

Now the last equation from the system (7)–(10) takes the form

$$\sum_{i=1}^{3} F_{i0}(\omega_i)\frac{\partial\omega_i}{\partial x_j}\frac{\partial\omega_i}{\partial x_j} + T_0(t) + \sum_{i=1}^{3} l_i = 0.$$

Splitting this relation with respect to independent variables $\omega_1, \omega_2, \omega_3, t$ for each class of functions $\vec{z} = \vec{z}(\vec{\omega})$ given in (13) yields the explicit forms of the functions $F_{01}(\omega_1), F_{02}(\omega_2), F_{03}(\omega_3)$ and $T_0(t)$ up to the choice of $\lambda_i, i = 1, 2, 3$ in (6)

$$F_{i0} = 0, \qquad T_0 = -\sum_{i=1}^{3} l_i. \tag{32}$$

We summarize the above-obtained results in the form of the following assertion.

**Theorem 1** *The Fokker-Planck equation (1) admits separation of variables if the drift coefficients $\vec{B}(\vec{x})$ are linear and given by formulae (20), where the matrix $\mathcal{M}$ is defined by formulae (19), (26), (27), (30) and (31).*

The coordinate systems allowing for variable separation in the corresponding FPE are given implicitly by formulae (11), (13) and (14), where $\mathcal{T}(t)$ is given in (26), (27) and (31), functions $h_i(t)$, $i = 1, 2, 3$ are given in (30) and functions $w_i(t)$, $i = 1, 2, 3$ are solutions of system of ordinary differential equations (22). Further details on explicit forms of the drift coefficients and the coordinate systems are given in Section III.

## III. Exact solutions

Remarkably, for the equation under study it is possible to give a complete account of solutions with separated variables. They have the form (2) and the separation equations for the functions $\varphi_\mu$, ($\mu = 0, 1, 2, 3$) read as (6), where the coefficients $F_{ai}$, $a, i = 1, 2, 3$ are the entries of the corresponding Stäckel matrices (15), functions $T_a$, $a = 1, 2, 3$ are listed in (16) and the functions $T_0, F_{a0}, a = 1, 2, 3$ given in (32).

The separation equation for the function $\varphi_0(t)$ is easily integrated. The separation equations for the functions $\varphi_i(\omega_i)$, $(i = 1, 2, 3)$ are similar to those arising from separation of variables in the Helmholtz equation $(\Delta_3 + \omega^2)\Psi = 0$. The solutions of these equations are well known (see, [6, 13] and the references therein). Below we adduce solutions of FPE (1) for each class of functions $\vec{z} = \vec{z}(\vec{\omega})$ given in (13).

1. Cartesian coordinates

$$u(t,\vec{\omega}) = \exp\left\{\sum_{i=1}^{3}\left(\lambda_i \frac{c_i^{-2}}{2l_i}e^{-2l_i t} - l_i t\right)\right\} \exp(i(\alpha\omega_1 + \beta\omega_2 + \gamma\omega_3))$$

and $\lambda_1 = -\alpha^2, \lambda_2 = -\beta^2, \lambda_3 = -\gamma^2$.

2. Cylindrical coordinates

$$u(t,\vec{\omega}) = \exp\left\{\lambda_1 \frac{c_1^{-2}}{2l_1}e^{-2l_1 t} + \lambda_3 \frac{c_3^{-2}}{2l_3}e^{-2l_3 t} - (2l_1 + l_3)t\right\} \times$$
$$J_n(\alpha e^{\omega_1}) \exp(i(n\omega_2 + \gamma\omega_3)),$$

where $J_n$ is the Bessel function [16, 14], and $\lambda_1 = -\alpha^2, \lambda_2 = -n^2, \lambda_3 = -\gamma^2$.

3. Parabolic cylindrical coordinates

$$u(t,\vec{\omega}) = \exp\left\{\lambda_1 \frac{c_1^{-2}}{2l_1}e^{-2l_1 t} + \lambda_3 \frac{c_3^{-2}}{2l_3}e^{-2l_3 t} - (2l_1 + l_3)t\right\} \times$$
$$D_{i\mu-1/2}(\pm\sigma\omega_1) D_{-i\mu-1/2}(\pm\sigma\omega_2) e^{i\omega_3\gamma},$$

where $\sigma = e^{i\pi/4}(2\alpha)^{1/2}$, $D_\nu$ is the parabolic cylinder function [17, 14] and $\lambda_1 = -\alpha^2, \lambda_2 = -2\alpha\mu, \lambda_3 = -\gamma^2$.

4. For the case of elliptic cylindrical coordinates we have two types of solutions

$$u(t,\vec{\omega}) = \exp\left\{\lambda_1 \frac{c_1^{-2}}{2l_1}e^{-2l_1 t} + \lambda_3 \frac{c_3^{-2}}{2l_3}e^{-2l_3 t} - (2l_1 + l_3)t\right\} \times$$
$$\text{Ce}_n(\omega_1, q)\,\text{ce}_n(\omega_2, q)\,e^{i\omega_3\gamma}, \quad n = 0, 1, 2, \ldots,$$
$$u(t,\vec{\omega}) = \exp\left\{\lambda_1 \frac{c_1^{-2}}{2l_1}e^{-2l_1 t} + \lambda_3 \frac{c_3^{-2}}{2l_3}e^{-2l_3 t} - (2l_1 + l_3)t\right\} \times$$
$$\text{Se}_n(\omega_1, q)\,\text{se}_n(\omega_2, q)\,e^{i\omega_3\gamma}, \quad n = 1, 2, 3, \ldots,$$

where $\text{ce}_n, \text{se}_n$ are the even and odd Mathieu functions, $\text{Ce}_n, \text{Se}_n$ are the even and odd modified Mathieu functions [14, 18] and $\lambda_1 = -4qa^2, \lambda_2 = 2q + c_n, \lambda_3 = -\gamma^2$, and $c_n$ are eigenvalues of the Mathieu functions.

5. Spherical coordinates

$$u(t,\vec{\omega}) = \exp\left\{\lambda_1 \frac{c_1^{-2}}{2l_1} e^{-2l_1 t} - 3l_1 t\right\} \times$$
$$\omega_1^{1/2} J_{\pm(n+1/2)}(\alpha/\omega_1) P_n^{\pm m}(\tanh\omega_2) e^{i\omega_3 m},$$

where $J_\nu$ is the Bessel function, $P_n^m$ is the Legendre function [14] and $\lambda_1 = -\alpha^2, \lambda_2 = -n(n+1), \lambda_3 = -m^2$.

6. Prolate spheroidal coordinates

$$u(t,\vec{\omega}) = \exp\left\{\lambda_1 \frac{c_1^{-2}}{2l_1} e^{-2l_1 t} - 3l_1 t\right\} \times$$
$$\operatorname{Ps}_n^{|m|}(\coth\omega_1, -a^2\lambda_1) \operatorname{Ps}_n^{|m|}(\tanh\omega_2, -a^2\lambda_1) e^{im\omega_3},$$

where $m$ is integer, $n = 0, 1, 2, \ldots, -n \le m \le n$, $\operatorname{Ps}_n^m$ is the spheroidal wave function [18] and $\lambda_2 = \lambda_n^{|m|}, \lambda_3 = -m^2$.

7. Oblate spheroidal coordinates

$$u(t,\vec{\omega}) = \exp\left\{\lambda_1 \frac{c_1^{-2}}{2l_1} e^{-2l_1 t} - 3l_1 t\right\} \times$$
$$\operatorname{Ps}_n^{|m|}(-i\tan\omega_1, -a^2\lambda_1) \operatorname{Ps}_n^{|m|}(\tanh\omega_2, a^2\lambda_1) e^{im\omega_3},$$

where $m$ is integer, $n = 0, 1, 2, \ldots, -n \le m \le n$, $\operatorname{Ps}_n^m$ is the spheroidal wave function and $\lambda_2 = \lambda_n^{|m|}, \lambda_3 = -m^2$.

8. Parabolic coordinates

$$u(t,\vec{\omega}) = \exp\left\{\lambda_1 \frac{c_1^{-2}}{2l_1} e^{-2l_1 t} - 3l_1 t\right\} e^{im\omega_3} \times$$
$$e^{m\omega_1} \exp(\pm i\alpha e^{2\omega_1}/2) \,_1F_1(-i\lambda_2/4\alpha + (m+1)/2, m+1, \mp i\alpha e^{2\omega_1}) \times$$
$$e^{m\omega_2} \exp(\pm i\alpha e^{2\omega_2}/2) \,_1F_1(i\lambda_2/4\alpha + (m+1)/2, m+1, \mp i\alpha e^{2\omega_2}),$$

where $_1F_1$ is the confluent hypergeometric function [17, 14] and $\lambda_1 = -\alpha^2, \lambda_3 = -m^2$.

9. Paraboloidal coordinates

$$u(t,\vec{\omega}) = \exp\left\{\lambda_1 \frac{c_1^{-2}}{2l_1} e^{-2l_1 t} - 3l_1 t\right\} \mathrm{gc}_n(i\omega_1; 2a\alpha, \lambda_2/2\alpha) \times$$
$$\mathrm{gc}_n(\omega_2; 2a\alpha, \lambda_2/2\alpha)\,\mathrm{gc}_n(i\omega_3 + \pi/2; 2a\alpha, \lambda_2/2\alpha)$$

or the same form with $\mathrm{gc}_n$ replaced by $\mathrm{gs}_n$. Here $\mathrm{gc}_n$ and $\mathrm{gs}_n$ are the even and odd nonpolynomial solutions of the Whittaker-Hill equation [19] and $n = 0, 1, 2, \ldots$, and what is more, $\lambda_1 = -\alpha^2$, $\lambda_3 = \mu_n$.

10. Ellipsoidal coordinates

$$u(t,\vec{\omega}) = \exp\left\{\lambda_1 \frac{c_1^{-2}}{2l_1} e^{-2l_1 t} - 3l_1 t\right\} \mathrm{el}_n^m(\omega_1)\,\mathrm{el}_n^m(\omega_2)\,\mathrm{el}_n^m(\omega_3),$$

where $m$ is integer, $n = 0, 1, 2, \ldots$, $-n \leq m \leq n$, $\mathrm{el}_n^m$ is the ellipsoidal wave function [18] and $\lambda_1 = \nu_{nm}$, $\lambda_2 = \lambda_{nm}$, $\lambda_3 = \mu_{nm}$.

11. For the case of conical coordinates we have two types of solutions

$$u(t,\vec{\omega}) = \exp\left\{\lambda_1 \frac{c_1^{-2}}{2l_1} e^{-2l_1 t} - 3l_1 t\right\} \omega_1^{\frac{1}{2}} J_{\pm(n+\frac{1}{2})}(\alpha/\omega_1) \times$$
$$\mathrm{Ec}_n^m(\omega_2)\,\mathrm{Ec}_n^m(\omega_3), \quad n = 0, 1, 2, \ldots, \quad m = 0, 1, \ldots, n,$$
$$u(t,\vec{\omega}) = \exp\left\{\lambda_1 \frac{c_1^{-2}}{2l_1} e^{-2l_1 t} - 3l_1 t\right\} \omega_1^{\frac{1}{2}} J_{\pm(n+\frac{1}{2})}(\alpha/\omega_1) \times$$
$$\mathrm{Es}_n^m(\omega_2)\,\mathrm{Es}_n^m(\omega_3), \quad n = 1, 2, 3, \ldots, \quad m = 1, 2, \ldots, n,$$

where $J_\nu$ is the Bessel function, $\mathrm{Ec}_n^m$ and $\mathrm{Es}_n^m$ are the even and odd Lamé functions [14, 18] and $\lambda_1 = -\alpha^2$, $\lambda_2 = -n(n+1)$, $\lambda_3 = -c_n^m$, where $c_n^m$ are eigenvalues of the Lamé functions.

In these equations we suppose that $l_i \neq 0$, $(i = 1, 2, 3)$. Given the condition $l_i = 0$, the expressions $\exp(-2l_i t)/2l_i$ should be replaced by $-t$.

Finally, we give a list of the drift velocity vectors $\vec{B}(\vec{x})$ providing separability of the corresponding FPEs. They have the following form:

$$\vec{B}(\vec{x}) = \mathcal{M}\vec{x} + \vec{v},$$

where $\vec{v}$ is arbitrary constant vector and $\mathcal{M}$ is constant matrix given by one of the following formulae:

1. $\mathcal{M} = \mathcal{T}L\mathcal{T}^{-1}$, where

$$L = \begin{pmatrix} l_1 & 0 & 0 \\ 0 & l_2 & 0 \\ 0 & 0 & l_3 \end{pmatrix},$$

$l_1, l_2, l_3$ are constants and $\mathcal{T}$ is an arbitrary constant $3 \times 3$ orthogonal matrix, i.e. $\mathcal{M}$ is a real symmetric matrix with eigenvalues $l_1, l_2, l_3$.

(a) $l_1, l_2, l_3$ are all distinct. The corresponding FPE has solution 1 only from the above list. The new coordinates $\omega_1, \omega_2, \omega_3$ are given implicitly by formula

$$\vec{x} = \mathcal{T} H(t)\, \vec{z}(\vec{\omega}) + \vec{w}(t), \tag{33}$$

where $\vec{z}(\vec{\omega})$ is given by formula 1 from (13), $\vec{w}(t)$ is solution of system of ordinary differential equations (22) and

$$H(t) = \begin{pmatrix} c_1 e^{l_1 t} & 0 & 0 \\ 0 & c_2 e^{l_2 t} & 0 \\ 0 & 0 & c_3 e^{l_3 t} \end{pmatrix} \tag{34}$$

with arbitrary constants $c_1, c_2, c_3$.

(b) $l_1 = l_2 \neq l_3$. The corresponding FPE has solutions 1–4 only from the above list. The new coordinates $\omega_1, \omega_2, \omega_3$ are given implicitly by (33), where $\vec{z}(\vec{\omega})$ is given by one of the formulae 1–4 from (13) and $H(t)$ is given by (34) with arbitrary constant $c_1, c_2, c_3$ satisfying the condition $c_1 = c_2$ for the partially split coordinates 2–4 from (13).

(c) $l_1 = l_2 = l_3$, i.e. $M = l_1 I$, where $I$ is unit matrix. The corresponding FPE has all 11 solutions, listed above. The new coordinates $\omega_1, \omega_2, \omega_3$ are given implicitly by formula (33), where $\vec{z}(\vec{\omega})$ is given by one of the eleven formulae (13) and $H(t)$ is given by (34) with arbitrary constants $c_1, c_2, c_3$ satisfying the condition $c_1 = c_2$ for the partially split coordinates 2–4 from (13) and the condition $c_1 = c_2 = c_3$ for the non-split coordinates 5–11 from (13).

2.
$$M = b\, \mathcal{C}_1 \begin{pmatrix} 0 & \cos s & 0 \\ -\cos s & 0 & \sin s \\ 0 & -\sin s & 0 \end{pmatrix} \mathcal{C}_1^{-1} + l_1 I,$$

where $I$ is the unit matrix and $\mathcal{C}_1$ is an arbitrary constant $3\times 3$ orthogonal matrix, $b, s, l_1$ are arbitrary constants and $b \neq 0$. The corresponding FPE has all 11 solutions, listed above with $l_1 = l_2 = l_3$. The new coordinates $\omega_1, \omega_2, \omega_3$ are given implicitly by formula (11), where $\vec{z}(\vec{\omega})$ is given by one of the eleven formulae (13), $\mathcal{T}(t)$ is given by (26)–(27), $\vec{w}(t)$ is solution of system of ordinary differential equations (22) and

$$H(t) = \exp(l_1 t) \begin{pmatrix} c_1 & 0 & 0 \\ 0 & c_2 & 0 \\ 0 & 0 & c_3 \end{pmatrix}$$

with arbitrary constants $c_1, c_2, c_3$ satisfying the condition $c_1 = c_2$ for the partially split coordinates 2–4 from (13) and the condition $c_1 = c_2 = c_3$ for non-split coordinates 5–11 from (13).

3.
$$M = \mathcal{C}_1 \begin{pmatrix} \frac{1}{2}(l_1 + l_3 + (l_1 - l_3)\cos 2s) & b\cos s & \frac{1}{2}(l_3 - l_1)\sin 2s \\ -b\cos s & l_1 & b\sin s \\ \frac{1}{2}(l_3 - l_1)\sin 2s & -b\sin s & \frac{1}{2}(l_1 + l_3 - (l_1 - l_3)\cos 2s) \end{pmatrix} \mathcal{C}_1^{-1},$$

where $\mathcal{C}_1$ is an arbitrary constant $3 \times 3$ orthogonal matrix, $b, s, l_1, l_2$ are arbitrary constants, $l_1 \neq l_3$ and $b \neq 0$. The corresponding FPE has solutions 1–4 only from the above list with $l_1 = l_2 \neq l_3$. The new coordinates $\omega_1, \omega_2, \omega_3$ are given implicitly by formula (11), where $\vec{z}(\vec{\omega})$ is given by one of the formulae 1–4 from (13), $\mathcal{T}(t)$ is given by (26), (27) and $(iii)$ from (31), $\vec{w}(t)$ is solution of system of ordinary differential equations (22) and

$$H(t) = \begin{pmatrix} c_1 e^{l_1 t} & 0 & 0 \\ 0 & c_2 e^{l_1 t} & 0 \\ 0 & 0 & c_3 e^{l_3 t} \end{pmatrix}$$

with arbitrary constants $c_1, c_2, c_3$ satisfying the condition $c_1 = c_2$ for the partially split coordinates 2–4 from (13).

Note that the above obtained solutions can be used as the basis functions to expand an arbitrary smooth solution of the equation under study in a properly chosen Hilbert space (for more details, see [6]).

The physical analysis of the obtained results seems to be very interesting, but the detailed study of this problem goes beyond the scope of the present paper.

## IV. R–separation of variables in the Fokker–Planck equation

In this paper, we restrict ourselves to the choice of separation Ansatz in the form (2). Generally speaking, the problem of separation of variables includes the search of *R–separable* solutions of the more general form [6]

$$u(t, \vec{x}) = e^{R(t,\vec{x})} \varphi_0(t) \prod_{a=1}^{3} \varphi_a \left( \omega_a(t, \vec{x}), \vec{\lambda} \right). \qquad (35)$$

In this case we have an analog of the system of equations (9)–(10)

$$\left( 2\frac{\partial R}{\partial x_j} + B_j \right) \frac{\partial \omega_a}{\partial x_j} + \frac{\partial \omega_a}{\partial t} + \Delta \omega_a = 0, \quad a = 1, 2, 3; \qquad (36)$$

$$\sum_{i=1}^{3} F_{i0}(\omega_i) \frac{\partial \omega_i}{\partial x_j} \frac{\partial \omega_i}{\partial x_j} + \frac{\partial R}{\partial t} + \Delta R + B_a \frac{\partial R}{\partial x_a} +$$

$$+ \frac{\partial R}{\partial x_a} \frac{\partial R}{\partial x_a} + T_0(t) + \frac{\partial B_a}{\partial x_a} = 0. \qquad (37)$$

Equations (7)–(8) are not changed. In a way analogous to that used above we get from (36) the form of the drift coefficients $\vec{B}(\vec{x})$

$$\vec{B}(\vec{x}) = \mathcal{M}(t)(\vec{x} - \vec{w}) + \dot{\vec{w}} - 2\vec{\nabla} R, \qquad (38)$$

where $\mathcal{M}(t)$ is given by formula (19).

The compatibility conditions of the above system of PDEs (38) yield

$$\begin{aligned}
B_{1x_2} - B_{2x_1} &= -2(\dot\alpha + \dot\beta \cos\gamma), \\
B_{1x_3} - B_{3x_1} &= -2(\dot\beta \cos\alpha \sin\gamma - \dot\gamma \sin\alpha), \\
B_{2x_3} - B_{3x_2} &= -2(\dot\beta \sin\alpha \sin\gamma + \dot\gamma \cos\alpha).
\end{aligned} \qquad (39)$$

As the functions $B_1, B_2, B_3$ are independent of $t$, it follows from these conditions that $\text{rot}\vec{B} = \text{const}$ and the functions $\alpha(t), \beta(t), \gamma(t)$ obey the system of ODE (25). Thus the matrix $\mathcal{T}(t)$ have the form (26).

Consequently the following assertion holds true.

**Theorem 2** *For the Fokker-Planck equation (1) to be R–separable it is necessary that the rotor of the drift velocity vector $\vec{B}(\vec{x})$ is constant.*

## Concluding Remarks

It follows from Theorem 1 that the choice of the drift coefficients $\vec{B}(\vec{x})$ allowing for variable separation in the corresponding FPE is very restricted. Namely, they should be linear in the spatial variables $x_1, x_2, x_3$ in order to provide separability of FPE (1) into three second-order ordinary differential equations. However, if we allow for separation equations to be of lower order, then additional possibilities for variable separation in FPE arise. As an example, we give the drift coefficients

$$B_1(\vec{x}) = 0, \quad B_2(\vec{x}) = 0, \quad B_3(\vec{x}) = B_3\left(\sqrt{x_1^2 + x_2^2}\right),$$

where $B_3$ is arbitrary smooth function. FPE (1) with these drift coefficients separates in the cylindrical coordinate system $t, \omega_1 = \ln\left(\sqrt{x_1^2 + x_2^2}\right), \omega_2 = \arctan(x_1/x_2), \omega_3 = x_3$ into two first-order and one second-order ordinary differential equations.

For the one-dimensional FPE the choice of the drift coefficients $\vec{B}(\vec{x})$ allowing for variable separation is essentially wider [4].

## Acknowledgement

I would like to thank Renat Zhdanov for his time, suggestions, and encouragement.